\documentclass[aps,twocolumn,reprint,superscriptaddress]{revtex4-1}
\usepackage{amsmath}
\usepackage{epsfig}
\usepackage{graphicx}
\usepackage{graphicx, color, epstopdf}
\usepackage{bm}
\usepackage{amssymb}
\usepackage[colorlinks, linkcolor=red,
anchorcolor=green,citecolor=blue ]{hyperref}
\usepackage{xcolor}
\usepackage{subfigure}

\begin{document}

\title{Uncovering quantum characteristics of incipient evolutions at
the photosynthetic oxygen evolving complex}
\author{Pei-Ying Huo}
\affiliation{School of Physics, Southeast University, Nanjing
211189, China}

\author{Wei-Zhou Jiang}
\email{wzjiang@seu.edu.cn}
\affiliation{School of Physics, Southeast University, Nanjing
211189, China}
\author{Rong-Yao Yang}
\affiliation{School of Physics, Southeast University, Nanjing
211189, China}

\author{Xiu-Rong Zhang}
\affiliation{School of Science, Jiangsu University of Science and
Technology, Zhenjiang 212100, China}

\begin{abstract}
Water oxidation of photosynthesis at the oxygen evolving complex
(OEC) is driven by  the polarization field induced by the
photoelectric hole. By highlighting the role of the polarization
field  in reshaping the spin and orbit potentials, we reveal in this work the characteristics and
underlying mechanism in   the relatively simpler OEC evolutions
within the states $\rm S_0\sim \rm S_2$   prior to the water
oxidation. The characteristic shifts of the density of states
(DOS) of the electron donor Mn atom are observed in the vicinity
of the Fermi surface to occur with the spin flips of Mn atoms and
the change of the Mn oxidation states during the electron
transfer.  Notably,  the  spin flips of Mn atoms point to the
resulting spin configuration of the next states. It is found that
the electron transfer tend to stabilize the catalyst OEC itself,
whereas the proton transfer pushes the evolution forward by
preparing a new electron donor, demonstrating the proton-coupled electron transfer. Meanwhile, it shows that the Mn-O bonds around the candidate Mn atom of the electron donor undergo
characteristic changes in the bond lengths during the electron
transfer. These concomitant phenomena uncovered in
first-principle calculations characterize the essential
equilibrium of the OEC between the state evolution and stability
that forms a ground  of the dynamic OEC cycles. In particular, the
characteristic undulation of the DOS around the Fermi level occurring at the
proton-coupled electron transfer can be used to  reveal  crucial processes in a wide range of realistic systems.
\end{abstract}

\maketitle

\section{Introduction}
The photosynthesis plays a crucial role in producing the biochemical
energy to sustain the Earth's biological cycle. In converting the
solar energy  into the biochemical energy, the photosystem  needs a
carousal of electrons eventually from the water oxidation. Water
oxidation is thus a fundamental reaction that is catalyzed at the
oxygen evolving complex (OEC) in photosystem II (PSII)
~\cite{shen2015structure,vinyard2017progress, chen2021bioinspired,
shamsipur2018latest, yocum2022photosystem}. The whole catalytic
process has long been thought to experience five redox reactions in
a closed cycle, dubbed Kok's cycle($\rm S_i$,
i=$0,\cdots,$4)~\cite{kok1970cooperation}, driven by the
photoelectric hole at the PSII reaction centre,
P$_{680}$~\cite{najafpour2020water}, whereas its complete
understanding remains  challenging and a hot frontier.

In recent decade, the striking advances have arisen mainly from the
measurements of the structures of $\rm S_i$ states using the
high-resolution X-ray diffraction of the crystalized photosystems
since the identification of the  S$_1$ state structure in
2011~\cite{umena2011crystal}. From then on,  a series of experiments
have progressively performed to identify the structures of S$_2$ and
S$_3$ states~\cite{pokhrel2014oxygen, cox2014electronic,
young2016structure, suga2017light, kern2018structures,
wilson2018structural,ibrahim2020, hussein2021structural}. Though the
structure of S$_4$ state remains elusive, latest detection using
serial femtosecond X-ray crystallography identified the presence of
this intermediate on the time sequence of the
evolution~\cite{bhowmick2023structural}. Another annual progress
involving the S$_4$ state and subsequent oxygen radical
identification was made strikingly upon the microsecond infrared
spectroscopy of various  vibration modes of protonated carboxylate
sidechains~\cite{greife2023electron}. The two different experiments
provide complementary information on the $\rm S_4$ state and the
mechanism of the $\rm O_2$ formation, but the characteristic time
scale of events remains rather different in these identifications.
In addition to the structures of five $\rm S_i$ states, more
information is indeed involved in the OEC reactions that include
more intermediates associated with redox processes, deprotonations,
water insertions and splitting, and the site-directed ionic
interactions as well~\cite{yocum2022photosystem,avramov2020role}.

For the supercomplex system that features multiple intermediates and
piecewise evolution processes, in pursuit of simplicity is a natural
strategy implemented not only in search of underlying laws but also
with appropriately chosen systems. Therefore,  this work focuses on
the relatively simple evolutions between $\rm S_0$ and $\rm S_2$
prior to the water insertion and dissociation. The underlying laws
for the evolutions of $\rm S_0\sim S_2$ can be used to understand
the more involved subsequent state  evolutions including the pending
$\rm S_4$ state. In particular,  the evolutions of
S$_0$$\rightarrow$S$_1$ and S$_3$$\rightarrow$S$_4$  are equally
characteristic of the transfer of an electron and a proton. The
state evolution is primely started by the photoelectric hole that
provides a Coulomb attraction to induce a serial electron transfer
through the tyrosine residue (Tyr161)
~\cite{wydrzynski2005photosystem, styring2012two,
barry1987tyrosine}. Concurrently, the polarization effect arising from the photoelectric hole and transferring electrons modifies the
the Coulomb potentials and consequently causes a series of  variations in electronic structures. In our previous work, it was found that the spin
flips of Mn atoms can serve as a marker under the polarization field
to trace the evolving intermediates and OEC
states~\cite{PhysRevApplied.21.024024}. In light of the spin flips,
this work aims to dig out systematically the characteristic signals
of the state evolutions under the polarization fields. In
particular, we will scrutinize the interplay between the
proton and electron transfers during the state evolutions in terms of the characteristic signals in the
electron density of the states (DOS).

In deed, the relationship between electron and proton transfers during the
state evolutions is scarcely elucidated in the literature and remains  largely unclear~\cite{shimizu2018mechanism,
dau2007eight, barry2015reaction, gagliardi2012role,
renger2004coupling, zaharieva2016sequential}. The depth of the recognition about the relationship is restrained by the limited evidences  that  the Tyr161 deprotonation is followed by  a backward proton migration cooperatively with the electron transfer from the CaMn$_4$O$_5$~\cite{nakamura2020pivotal,
nakamura2014fourier, styring2012two}, while the proton and electron transfer sequence  in the S$_2$ state seems to
rely on the initial open or closed geometry of the S$_2$ state~\cite{narzi2014pathway,
nakamura2020pivotal,PhysRevApplied.21.024024}. These situations lead naturally to the question
of  how  the proton and electron transfers couple  in the OEC  evolution. Notably, it is known that some redox in the aqueous or hydrogen-rich environments can occur with  the proton-coupled electron transfer (PCET).  Examples of the PCET range from chemical, catalytic to biological processes, such as  enzymatic C-H oxidation~\cite{mittra2019reduction, yosca2013iron, tyburski2021proton}, DNA synthesis and repair~\cite{stubbe2003radical, aubert2000intraprotein, jackson2019graphite}, ribonucleotide reduction~\cite{minnihan2013reversible},  and the interconversion reactions of   small molecules like O$_2$/H$_2$O~\cite{huynh2007proton, darcy2018continuum}, N$_2$/NH$_3$~\cite{tanabecatalytic, ashida2022catalytic}, and CO$_2$/alkanes~\cite{esmaeilirad2023imidazolium,  khezeli2024computational}. In this circumstance, broad significance can be assigned to our study on the characteristic relationship between the electron and proton transfers that is instructive for understanding a variety of  processes, say, in catalysis~\cite{costentin2020proton, tessensohn2019voltammetric}, synthetic chemistry~\cite{gentry2016synthetic, miller2016proton, murray2021photochemical} and energy conversion~\cite{hammes2009theory, nocera2022proton, goyal2017tuning}, and is of potential applications.

As the evolutions of $\rm S_0$ $\sim$ $\rm S_2$ are characterized
mainly by electron transfers in the redox reactions of  the
CaMn$_4$O$_5$ cluster, concomitant changes are weaved by the shift
of the local Coulomb field, the spin flips of Mn atoms, and
propagated wrinkles in the electron density of states (DOS) near the
Fermi surface. Using the density functional theory, we will search
for the characteristic phenomenologies among these changes  during
the  evolutions from S$_0$ to S$_2$ through $\rm S_1$. We will find
that the characteristic phenomenologies can serve as the precursor
to reveal the underlying physics of the water oxidation cycle.

\section{Computational Methods}
\label{method}
\begin{figure}[thb]
\centering
\includegraphics[width=0.6\columnwidth]{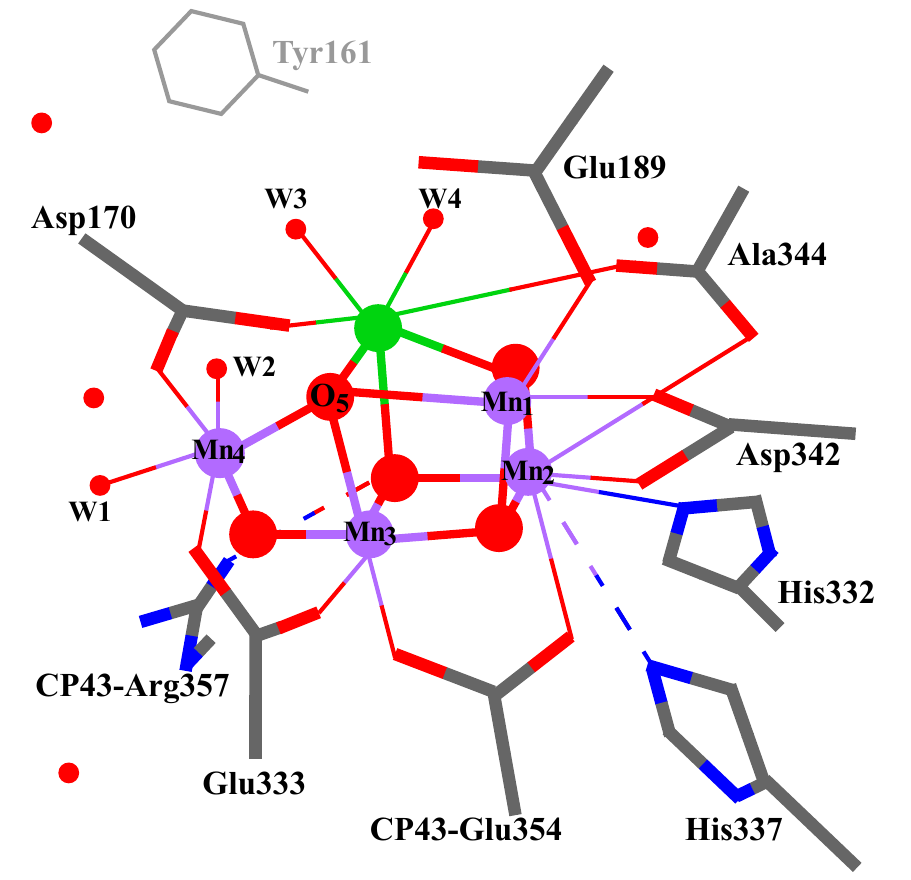}\\
\caption{Geometric model schematic diagram of the OEC CaMn$_4$O$_5$,
along with surrounding  ligated amino acid residues. Hydrogen
atoms are omitted for clarity, while the Tyr161 outside of the model is sketched. The Mn, Ca, and O atoms are represented by solid circles in purple, green, and red colors, respectively. The intersections of the grey, red, and blue solid lines correspond to the ligand C, O, and N atoms, respectively. Single red dots represent water molecules surrounding amino acid residue ligands.
}\label{fig1}
\end{figure}

We perform the spin-polarized all-electron density functional
simulations with the DMol$^3$ package in the Materials Studio of
Accelrys Inc.  The OEC system in the simulation is  a 112-atom
configuration whose schematic diagram in Fig.~\ref{fig1} is
intercepted from the XRD crystal data obtained at
1.9~\AA~resolution~\cite{umena2011crystal}. The prototype of the S$_1$ state is the 112-atom geometric model, and optimization is performed to obtain the S$_1$ state structure. The S$_0$ and S$_2$ states are the results of optimization, with the prototypes also arising from 112-atom geometric model of the S$_1$ state by adding back a hydrogen atom and removing an electron, respectively.

In the geometric
optimization process, we adopt the B3LYP hybrid
functional ~\cite{becke1992density, lee1988development,
vosko1980accurate, stephens1994ab}  and a double numerical basis set with polarization functions (DNP).
The Grimme method for dispersion correction (DFT-D) is applied to account for weak residue interactions. In the self-consistent field (SCF) process, the convergence criterion is set to 1.0$\times$10$^{-5}$ Ha, and the direct inversion in the iterative subspace (DIIS) method is used to accelerate SCF convergence. All eight possible Mn spin configurations($\uparrow\uparrow\uparrow\uparrow$, $\downarrow\uparrow\uparrow\uparrow$, $\uparrow\downarrow\uparrow\uparrow$, $\uparrow\uparrow\downarrow\uparrow$, $\uparrow\uparrow\uparrow\downarrow$, $\uparrow\downarrow\uparrow\downarrow$, $\uparrow\downarrow\downarrow\uparrow$, $\uparrow\uparrow\downarrow\downarrow$) of the CaMn$_4$O$_5$ structure are considered, and the configuration with the lowest energy is identified as the S$_\mathrm{i}$ state structure.

The dynamic process of the electron transfer to the hole at Tyr161
induced by the polarization field is simulated by setting a point
charge of the hole with 1.0e at Tyr161 and increasing the equivalent
charge of the transferring electron placed at the fixed distance of
2~\AA~ away from Ca atom of CaMn$_4$O$_5$ on the line connecting to
the O atom of Tyr161. The fractional equivalent charge such as
-0.1$\sim$-0.9e is placed to simulate the distance-dependent Coulomb
potential in a gradual transfer process of  the electron. The effect
of the position of the transferring electron is further verified by
moving the equivalent charge of -0.9e to the position  at a distance
of 4~\AA~ away from the Ca atom of  CaMn$_4$O$_5$,  which is closer
to the hole for further hole-electron recombination. With the Coulomb fields of the point charges being included, optimization calculations are performed. In the
simulation,  the constraint of the charge conservation is imposed on
the whole system. More details can be referred to
Ref.~\cite{PhysRevApplied.21.024024}. Here, modeling the electron transfer by setting  the fractional charge can be reasonably simulated by the ab initio molecular dynamics in a time scale of a few hundred of fs~\cite{narzi2014pathway,guo2024closing}.

\section{Results and Discussion}
\label{result}
The evolutions from S$_0$ to S$_2$ primarily involve the transfer of
electrons and protons from the CaMn$_4$O$_5$,  while the electron
transfer fulfills at the hole-electron recombination. The dynamic
electron transfer is associated with the variations in the bond
valence and polarization field. Hence, elucidating the dynamic
electron transfer can detail the animated evolution of
CaMn$_4$O$_5$ and extract the characteristic phenomena in the
presence of  the varying polarization field. For S$_0$$\sim$S$_2$
states,  the oxidation states of Mn$_1$$\sim$Mn$_4$ are reproduced
to be S$_0$ (III, IV, III, III), S$_1$ (III, IV, IV, III),
S$_2$-open (III, IV, IV, IV), and S$_2$-closed (IV, IV, IV, III),
with the spin configurations of Mn$_1$$\sim$Mn$_4$ being
$\uparrow\downarrow\uparrow\downarrow,
\uparrow\downarrow\downarrow\uparrow,
\uparrow\downarrow\downarrow\uparrow$, and
$\uparrow\uparrow\uparrow\downarrow$,
respectively~\cite{krewald2016spin, shen2015structure}.  Here, the spin configurations are verified to be consistent with those obtained from the complete active space self-consistent field calculation with the explicit inclusion of the spin-orbit coupling~\cite{retegan2016five}.

\begin{figure}[thb]
\centering
\includegraphics[width=0.9\columnwidth]{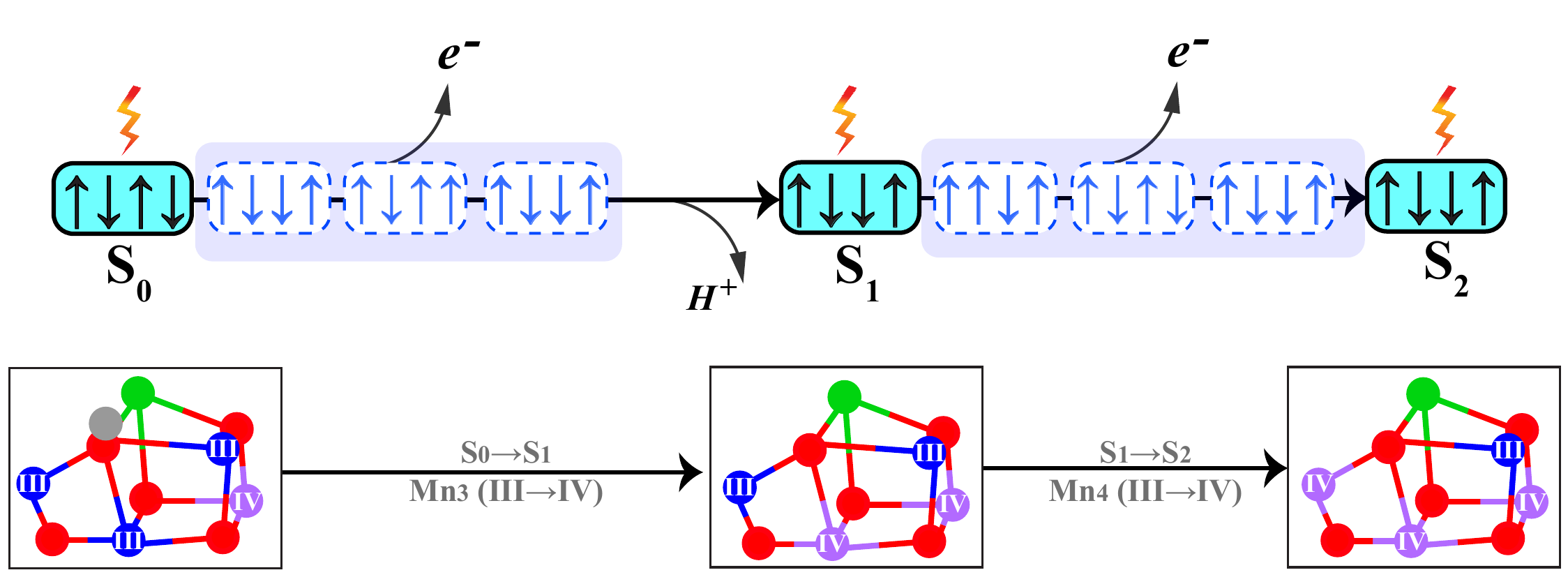}\\
\caption{
Evolutions of spin configurations and oxidation states of Mn atoms
for S$_0$$\sim$S$_2$. The spins in solid and dashed boxes are for
the $\rm S_i$ and intermediate states, respectively.  In the second
row, the Mn atoms are marked with oxidation states (III and IV), and
Ca, O, and H atoms are denoted by green, red, and grey spheres,
respectively.}
\label{fig2}
\end{figure}

When  the hole polarization field switches on  after the oxidation
of Tyr161,  the Mn spin configuration of the S$_0$ state flips from
$\uparrow\downarrow\uparrow\downarrow$ to
$\uparrow\downarrow\downarrow\uparrow$, accompanied by an increase
in the total energy of the system.  In response to the hole
polarization field, the system undergoes an excitation  to an
intermediate for subsequent electron transfer. The similar spin flip
of the S$_1$ state occurs in the hole polarization field with the Mn
spin configuration to $\uparrow\uparrow\downarrow\uparrow$. Then,
the candidate Mn atom of the electron donor loses one electron for
recombination with the hole at Tyr161.

In the process of the electron transfer from the OEC, the effective
charge of the transferring electron is set to vary from -0.1e to
-0.9e at a site 2 \AA~ away from the Ca atom of CaMn$_4$O$_5$ on the
line connecting to the O atom of Tyr161.  During this process, we
can observe that the spin configuration of Mn atoms undergoes some
intermediate configurations and flip exactly to that of the next S
state, see Fig.~\ref{fig2}. In the evolution from S$_0$ to S$_1$,
the Mn spin configuration ultimately changes to
$\uparrow\downarrow\downarrow\uparrow$ after going through
$\uparrow\downarrow\uparrow\uparrow$. When the released electron
moves closer to Tyr161 at a distance of 4~\AA~ away from the Ca atom
of the CaMn$_4$O$_5$, the  spin configuration of Mn atoms remains
unchanged. The subsequent proton removal also does not alter the Mn
spin configuration of  $\uparrow\downarrow\downarrow\uparrow$, which
then becomes the spin configuration of the S$_1$ state.  Similar
spin flips are observed in the process of an electron transfer from
the OEC of the $\rm S_1$ state, as shown in Fig.~\ref{fig2}. After
the electron transfer, the spin configuration shifts to that of the
$\rm S_2$-open. Since the occurrence of the spin flips of Mn atoms  is attributed to the electron transfer and the resulting shift in the Coulomb field induced by the hole polarization, the spin flips can serve as an
indicative  marker for the pivotal OEC evolution driven by the hole
polarization.

\begin{figure}[thb]
\centering
\includegraphics[width=0.97\columnwidth]{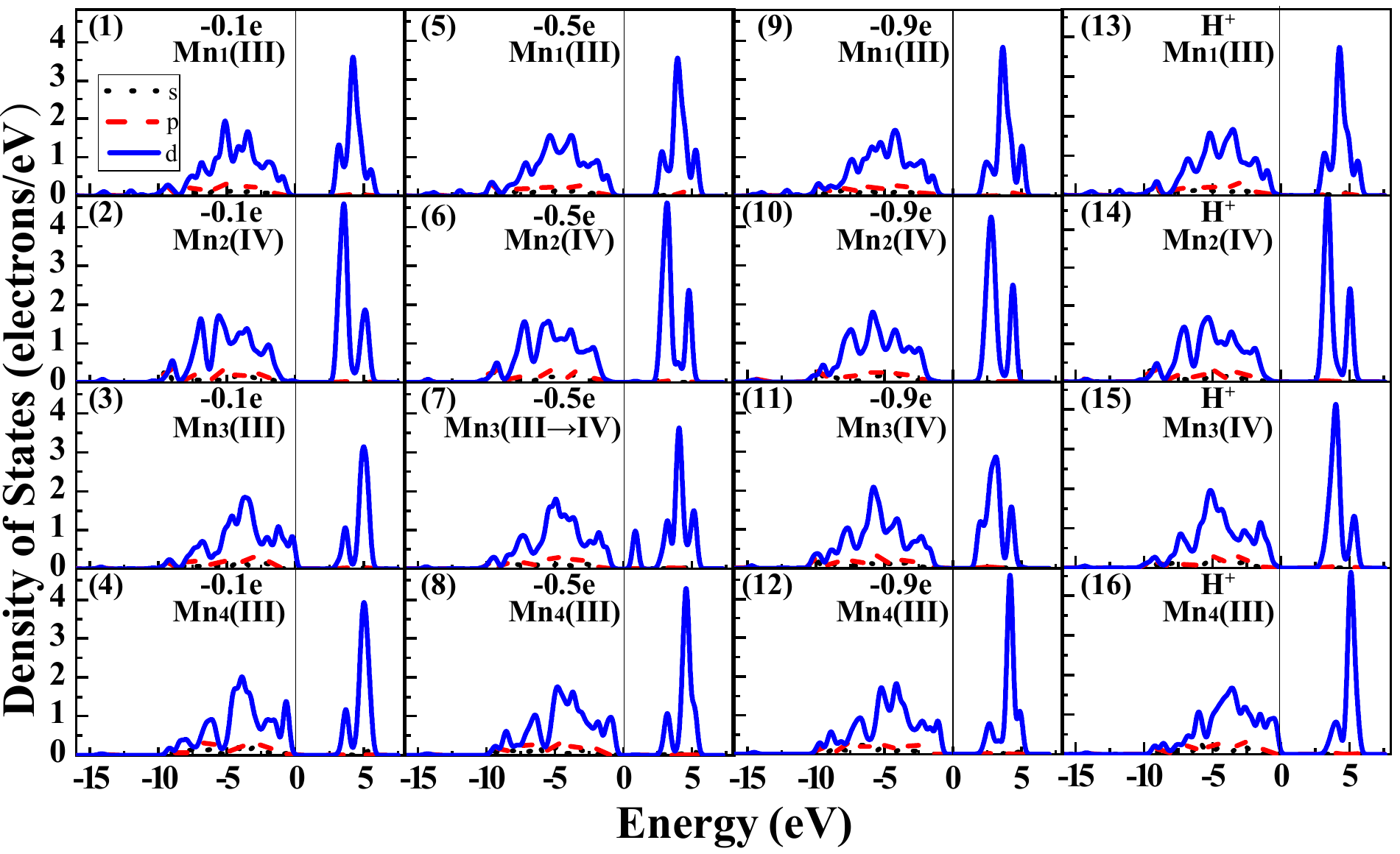}\\
\caption{
The PDOS of  s, p and d orbitals of Mn$_1$$\sim$Mn$_4$ during
S$_0$$\rightarrow$S$_1$. Column panels are with effective charge of
-0.1, -0.5 and -0.9 e of the transferring electron and with the
proton transfer, respectively. The vertical line represents the
Fermi surface.
}\label{fig3}
\end{figure}
Shown in Fig.~\ref{fig2} is also the variation in the bond valence
of Mn atom that represents for an electron transmission from below
the Fermi surface to the liberation of the Coulomb binding. This
demonstrates the necessity to examine the variation in the density
of states (DOS) especially near the Fermi surface in the process of
the electron transfer.   Figure~\ref{fig3} shows the variation in
the partial DOS (PDOS) of s, p and d orbitals for the atoms of
Mn$_1$$\sim$Mn$_4$ with increasing the effective charge of the
transferring electron at the specified position during the evolution
from S$_0$ to S$_1$. It is found that in the S$_0$ state the value
of the Mn$_3$ DOS of the d orbital  is significant at the site
sufficiently near Fermi surface.   We can observe from the third row
of Fig.~\ref{fig3} that the Mn$_3$ DOS of the d orbital comes across
the Fermi surface to form a small separate peak that blends into the
bumps above the Fermi surface in a process  where the electron
transfer is simulated by placing the effective charge of -0.1, -0.5
and -0.9e at the site specified above, respectively. Compared with
the sluggish changes in the PDOS of other Mn atoms, it clearly
suggests that Mn$_3$ is  the electron donor during the evolution
from S$_0$ to S$_1$, which is actually in accord with the variation
of the Mn oxidation state. During the electron transfer until its
fulfillment,  the PDOS boundaries of Mn atoms below the Fermi
surface experience  more or less a contraction  inwards to keep the
electron distributions  better  bound. This leads to a temporary
stability of the OEC in  the dynamical OEC evolution prior to the
proton transfer. The proton transfer pushes the PDOS of Mn atoms
outwards and especially brings  the Mn$_4$ PDOS to scrape over the
Fermi surface. This gives the opportunity and vigor for another
ignition.

\begin{figure}[thb]
\centering
\includegraphics[width=0.97\columnwidth]{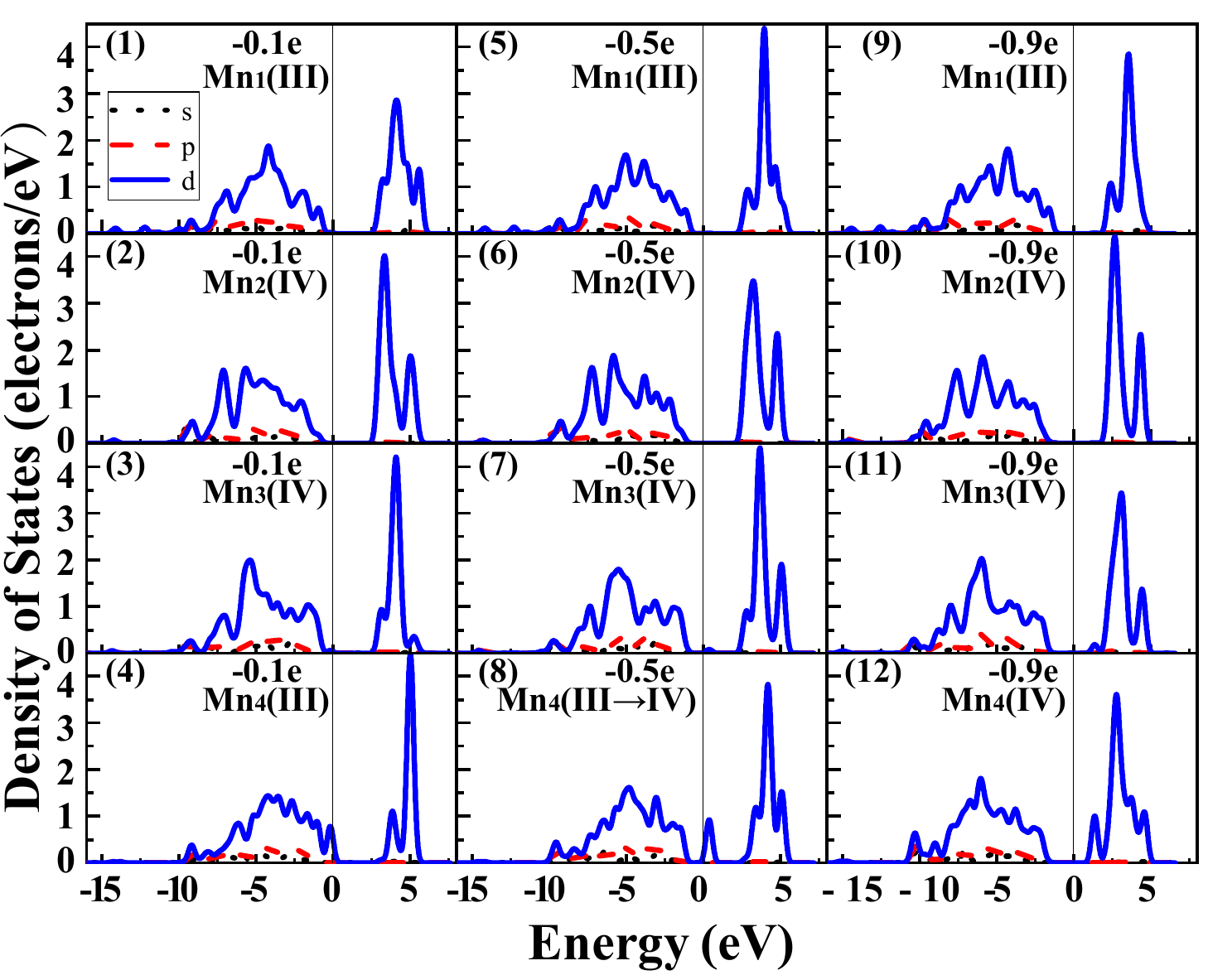}\\
\caption{
The same as in Fig.~\ref{fig3} but for S$_1$$\rightarrow$S$_2$
without proton transfer.
}\label{fig4}
\end{figure}

As another light flash breaks the tranquility, the photoelectric
hole is recreated to drive the evolution from the $\rm S_1$ to $\rm
S_2$ state. Similar variations in the Mn PDOS arise from the
electron transmission from below the Fermi surface. At this time, it
is the Mn$_4$ PDOS that shows the animation from coming across the
Fermi surface to undergoing the splitting and blending into
high-level distributions well above the Fermi surface, characterizing the Mn$_4$ oxidation as shown
clearly in Fig.~\ref{fig4}.
Similar to
the evolution from the S$_0$ to S$_1$ state, the release of electron
seems to stabilize the catalyst cluster.  This establishes the
essential equilibration  between the dynamic evolution and the state
stability that  forms a necessary foundation for the precessing OEC
cycle. Moreover, it should be stressed that the electron transfer
here is correlated with the preceding proton transfer with which the
hydrogen bonding network is associated
tightly~\cite{horke2016hydrogen}.

\begin{figure}[thb]
\centering
\includegraphics[width=0.9\columnwidth]{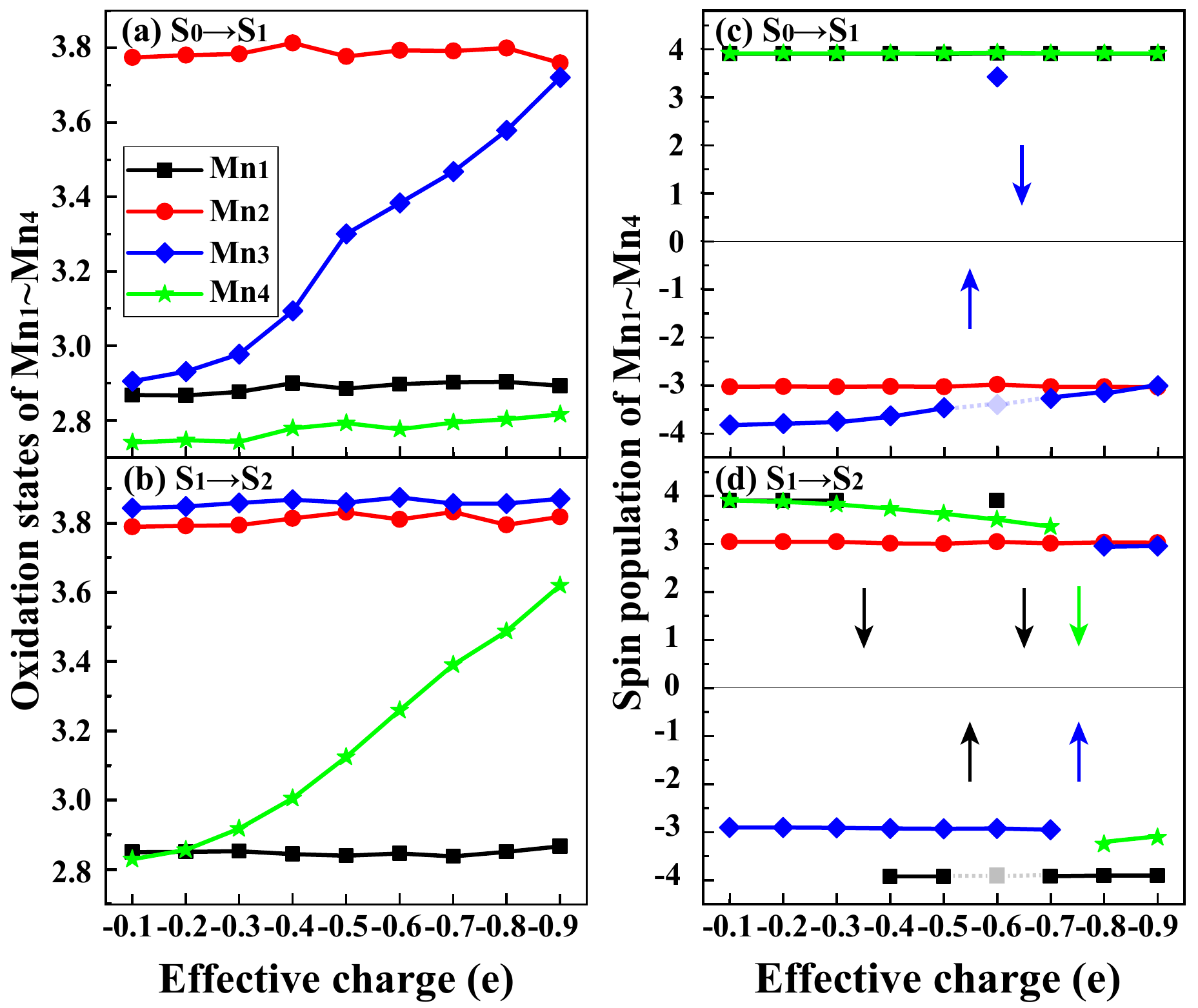}\\
\caption{
Variations of Mn$_1$$\sim$Mn$_4$ oxidation states (a) and (b), and
their Mulliken spin population (c) and  (d).  The effective charge
of the transferring electron varies from  $-0.1$ to $-0.9$ e with
the polarization field imposed. The arrows in (c) and (d) represent
the flip of spin orientation. }\label{fig5}
\end{figure}

During the electron transfer, the PDOS of all Mn atoms in the Fermi
sea is observed to undergo some rearrangements. As a result, the
concomitant  microscopic  changes are also found in the oxidation
states and spin configurations of Mn atoms, as shown in
Fig.~\ref{fig5}. Here, the Mn oxidation state is calculated from the
bond valence sum (BVS)  formula $ \rm S=\sum{exp(R_0-R_{ij})/B_0 }
$, where
R$_{ij}$ represents each bond length between the Mn and surrounding
O or N atoms, and other relevant parameters are taken as B$_0$=0.37,
R$_0$(Mn-O)=1.750~\AA, and R$_0$(Mn-N)=1.822~\AA~\cite{liu1993bond}.
It can be observed from Fig.~\ref{fig5}a and b that the variations
in the oxidation states of the candidate Mn atoms of the electron
donor are almost linearly correlated with the effective charge of
the transferring electron, while the oxidation states of other Mn
atoms remain nearly unchanged. Specifically, with the effective
charge of the transferring electron varying from $-0.1$ to $-0.9$ e,
the  Mn$_3$ oxidation state gradually changes from about 2.9 to 3.7
in S$_0$$\rightarrow$S$_1$ and Mn$_4$ changes from 2.8 to 3.7 in
S$_1$$\rightarrow$S$_2$, representing a valence change from III to
IV. During the electron transfer in the polarization field, the spin
flips of Mn atoms may occur with  the changes of Mn oxidation
states. Depicted in Fig.~\ref{fig5}c and d are the Mulliken spin
populations of Mn atoms that give the number of unpaired electrons
around Mn and the spin orientations. The positive and negative
values correspond to up and down spin orientations, respectively.
The reduction of the Mn$_3$ and Mn$_4$ spin populations from 4 to 3
in  S$_0$$\rightarrow$S$_1$ and S$_1$$\rightarrow$S$_2$ accords with
the shift of their oxidation states from III to IV for the electron
transfer. It can be seen from Fig.~\ref{fig5}c and d that some
intermediate states are produced  with the spin flips of Mn atoms
during the electron transfer. Especially, multiple flips of the Mn
spin orientations take place in the transfer process of
S$_1$$\rightarrow$S$_2$. With a succession of intermediate states in
the varying polarization field, the OEC states ($\rm S_0$, $\rm
S_1$) evolve to the next state with the electronic and geometric
structures consistent with those from the measurements and
theoretical computations~\cite{pal2013s0, siegbahn2013water,
koulougliotis1992oxygen, askerka2015nh3, pantazis2012two,
zimmermann1986electron, haddy2004q, astashkin1994pulsed}.

\begin{figure}[thb]
\centering
\includegraphics[width=0.9\columnwidth]{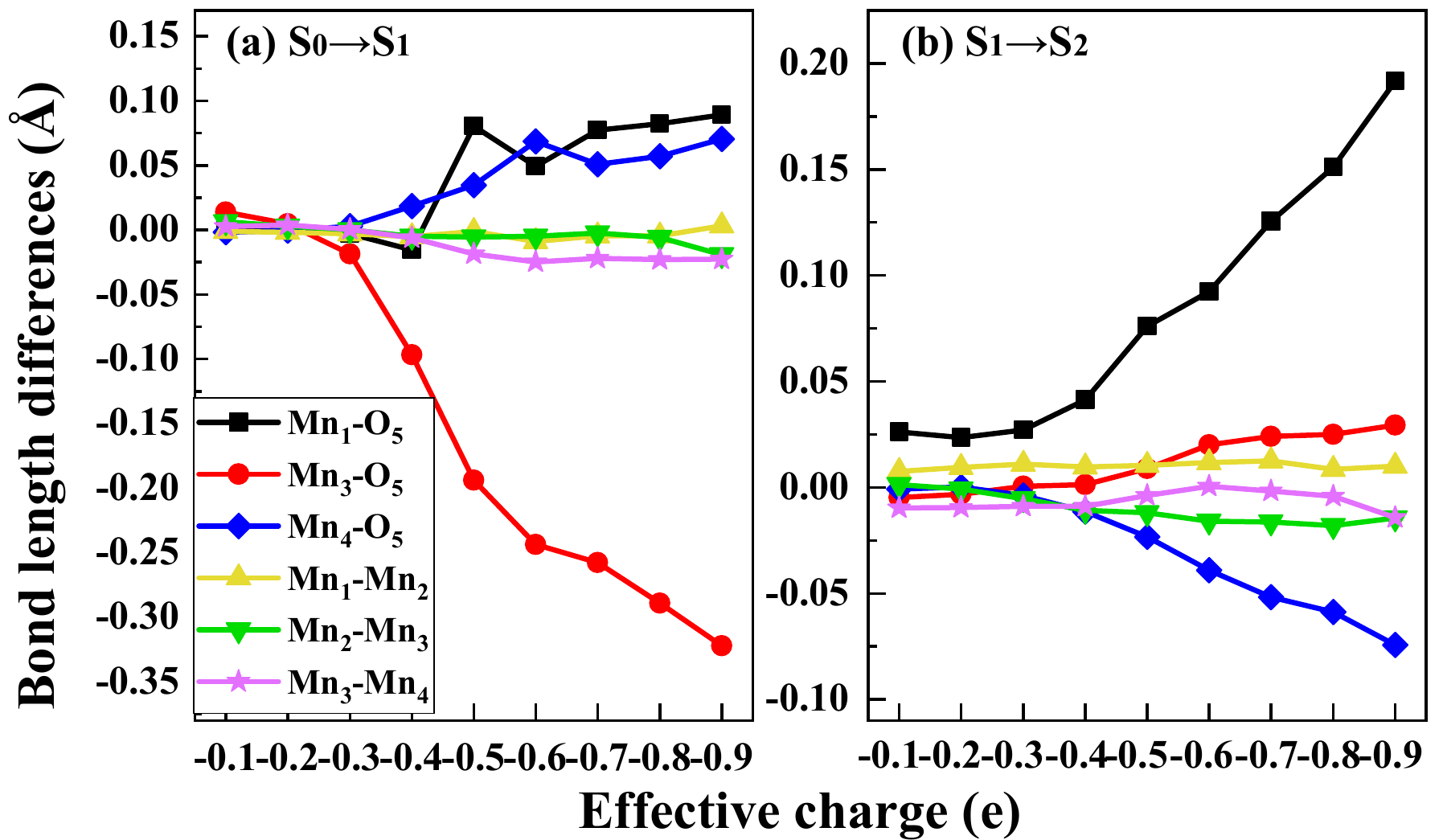}\\
\caption{
Variations of Mn-O$_5$ and Mn-Mn bond lengths. In (a) and (b), the
results are relative to those of the  S$_0$ and S$_1$ states,
respectively. The effective charge of the transferring electron
varies from $-0.1$ to $-0.9$ e with the polarization field imposed.
}\label{fig6}
\end{figure}

In the dynamical evolution, the above microscopic quantum shifts are
also accompanied  by the change in the phenomenological geometrical
structure.  Figure~\ref{fig6} displays the variations of Mn-O$_5$
and Mn-Mn bond length during the electron transfer under the
polarization field.  In Fig.~\ref{fig6}a and b, the variations of
the bond lengths during the electron transfer of
S$_0$$\rightarrow$S$_1$ and S$_1$$\rightarrow$S$_2$ are relative to
those  of  S$_0$  and S$_1$ states, respectively. Positive and
negative values stand for the elongation and shortening  of bond
lengths, respectively. It can be observed that the alteration in
Mn-O  bond lengths is more appealing due to the attraction between
them, while the less altered Mn-Mn distances suggest that the $\rm
S_i$ structures are rather stable during the electron transfer. In
Fig.~\ref{fig6}a, the main change is observed to be the considerable
shortening of the Mn$_3$-O$_5$ bond length that is  approximately
0.35~\AA~  with the effective charge of the transferring electron
varying from -0.1 to -0.9 e during the electron transfer from S$_0$
to S$_1$ state. The large shortening arises from the electron loss
of Mn$_3$ that gives rise to a stronger attraction between Mn$_3$
and O$_5$.

In the evolution of S$_1$$\rightarrow$S$_2$, the Mn$_1$-O$_5$ length
shows an overall trend of elongation, while the Mn$_4$-O$_5$ length
displays the opposite trend of shortening, as the effective charge
of the transferring electron varies from $-0.1$ to $-0.9$ e, see
Fig.\ref{fig5}b. The shortening of the Mn$_4$-O$_5$ bond length is
attributed to the enhanced attraction between Mn$_4$ and O$_5$ due
to the electron loss of Mn$_4$. In contrast, the  attraction between
Mn$_1$ and O$_5$ is weakened to cause the elongation of the bond
length.  Specifically, the Mn$_1$-O$_5$ length increases by about
0.2~\AA~ throughout the electron transfer process, corresponding to
the  Mn$_1$-O$_5$ bond length of 2.901 and 3.092~\AA~ in S$_1$ and
S$_2$-open state, respectively. The active variation of the
Mn$_1$-O$_5$ and Mn$_4$-O$_5$ bonds suggests the existence of the
live domain in the OEC structure. Actually, the counteraction
between the Mn$_1$-O$_5$ and Mn$_4$-O$_5$ bondings offers  support
for  the existence of  two structures that are characteristic of the
commodious space with the elongated bond length of Mn$_1$-O$_5$ and
Mn$_4$-O$_5$, called $\rm S_2$-open and $\rm
S_2$-closed~\cite{pantazis2012two}, respectively.

\section{Summary}
\label{summary}
In this work, we investigate the OEC evolutions during $\rm S_0\sim
S_ 2$ to reveal the animation and mechanisms therein by virtue of
the characteristic signals with the density functional simulations.
The electron transfer is eventually  driven by the hole
polarization, whereas a direct evidence of the correlation with the
proton transfer is revealed in the state evolution  during which the
electron donor Mn atom is readily prepared by the preceding proton
transfer.
The characteristic shift of the Mn PDOS near the Fermi surface is
found to correlate simultaneously with the variations in the Mn
oxidation states and spin flips, together with the change in the
geometric structure. These concurrent events consist of a collection
of the characteristic signals for the state evolution. In
particular,  the spin flips of Mn atoms,  consistent with variation
of the Mulliken spin populations,  can point exactly to the spin
configuration of the next $\rm S_i$ state after undergoing some
intermediates during electron transfer.  Importantly, the stability
of the OEC arising from the electron transfer brings the rhythm in
the theme of the dynamical evolution that is necessary for the
unceasing OEC cycle. Our findings  will paly an instructive role in
unveiling the pending $\rm S_4$ state and  revealing  more
underlying physics for the water oxidation cycle.
Moreover, the direct evidence of the correlation of the proton and electron transfers in the PDOS on the Fermi energy level can be instructively used as a tracer to tag and understand a wide range of biochemical processes and redox reactions that are characteristic of the PCET.

\section*{Acknowledgements}
 This work was supported in part by the National Natural Science
 Foundation of China under Grants No. 11775049 and No.
12375112. The Big Data Computing Center of Southeast University is
acknowledged for providing the facility support for the partial
numerical calculations of this work.

\end{document}